\documentclass{IEEEoj}
\usepackage{cite}
\usepackage{amsmath,amssymb,amsfonts}
\usepackage{algorithmic}
\usepackage{graphicx,color}
\usepackage{textcomp}

\usepackage{comment}
\usepackage[colorlinks=true, linkcolor=blue]{hyperref}

\def\equationautorefname~#1\null{Eq.~(#1)\null}
\usepackage{commath}

\def\BibTeX{{\rm B\kern-.05em{\sc i\kern-.025em b}\kern-.08em
    T\kern-.1667em\lower.7ex\hbox{E}\kern-.125emX}}
\AtBeginDocument{\definecolor{ojcolor}{cmyk}{0.93,0.59,0.15,0.02}}
\def\OJlogo{\vspace{-14pt}\includegraphics[height=28pt]{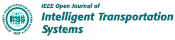}}
\begin{document}
\receiveddate{XX Month, XXXX}
\reviseddate{XX Month, XXXX}
\accepteddate{XX Month, XXXX}
\publisheddate{XX Month, XXXX}
\currentdate{XX Month, XXXX}
\doiinfo{OJITS.2022.1234567}

\title{Validation of a driver model for energy consumption simulations of road vehicles}

\author{~LUIGI ROMANO\authorrefmark{1}, 
MICHELE GODIO\authorrefmark{2}, 
FREDRIK BRUZELIUS\authorrefmark{3}, 
P\"{A}R JOHANNESSON\authorrefmark{2}}
\affil{Division of Vehicular Systems, Department of Electrical Engineering, Linköping University, SE-581 83 Linköping, Sweden}
\affil{RISE Research Institutes of Sweden, Department of Chemistry and Applied Mechanics, Brinellgatan 4, 504 62 Borås, Sweden}
\affil{Vehicle Engineering and Autonomous Systems, Department of Mechanical Engineering, Chalmers University of Technology, Chalmers University of Technology, Hörsalsvägen 7 A, SE-412 96 Göteborg, Sweden}
\corresp{CORRESPONDING AUTHOR: F. Bruzelius (e-mail: fredrik.bruzelius@chalmers.se).}
\authornote{This work was supported by the Swedish Energy Agency.}
\markboth{Preparation of Papers for IEEE OPEN JOURNALS}{Author \textit{et al.}}

\begin{abstract}
The energy performance of road vehicles has traditionally been evaluated using driving cycles, in which a prescribed speed profile is tracked either in simulation or by a physical vehicle. Recent research has proposed an alternative framework based on operating conditions, where the driving environment is described in terms of factors such as road topography, legal speed limits, traffic, and weather, rather than by an explicit speed profile. 
This paper addresses the complementary problem of driver modeling, necessary to translate the operating conditions to a speed profile.
A simple driver model designed to be compatible with the operating-condition framework and evaluates its ability to reproduce realistic driving behavior is investigated. Validation of the model is done through a controlled driving-simulator study vehicle log files collected during real-world operation and are used to identify the model parameters and assess the model's predictive performance. The results suggest that this simple model can be efficient in reproducing major effects of driver behavior, while further research is required to fully assess its validity.
\end{abstract}



\maketitle

\section{Introduction}
A driver model is a necessity to simulate complete vehicles. Applications of driver models range from microscopic traffic simulations in the car following models, see, e.g.,~\cite{Brackstone1999}, to ADAS assessment applications in near collisions as in \cite{markkula2012review} to fuel economy and emission simulations as in \cite{Mcgordon2011}. The purpose of a driver model also differs in and between applications. For some applications, it may serve as a tool to shift the natural causality in a simulation, see e.g.~\cite{pettersson2020intrinsic}, such as tracking a reference speed as good as possible or to assess the best possible performance achievable for a vehicle model as in \cite{Sharp2010}. Another important field of driver modeling is microscopic traffic simulation, where car simple following models of human/vehicle are used to study traffic phenomenon \cite{olstam2004comparison}. In other applications, the objective is to mimic human behavior for certain conditions and situations. Comprehensive reviews on driver models can be found in \cite{Nash2016}, as well as in \cite{Plochl2007} and \cite{Macadam2010}.

This paper is concerned with energy consumption simulations, and the objective of the driver model is to capture human behavior. The driving style and behavior of the driver have a documented influence on fuel consumption and emissions. In \cite{Walnum2015} a statistical analysis of logged data from the real operation of trucks in northern Sweden and Norway were analyzed w.r.t to energy consumption. It was concluded that a significant influence could be attributed to the differences in drivers, even though the attributes of the infrastructure rated a ten folded greater difference compared to the driver influence for these conditions. In \cite{Zheng2019} a similar analysis was carried out, but using a controlled experiment with participants in and between Chinese cities. The data set was categorized into four groups and a simple car following model was used to quantify the differences observed. The study in \cite{Evans1978} focused on traffic influence and the correlation between trip time and fuel consumption. It was concluded that drivers with longer experience could compensate for longer trip times and fuel consumption by planning their driving and keeping a more even speed. The study \cite{VanMierlo2004} used a combination of test bench measurements and traffic recordings to study traffic measures and driving styles' influence on emissions. A general conclusion is that traffic policies and Eco Driving companies that promote a more fluent way of driving have a great impact on the overall emissions in a region. Driving cycles designed to assess energy consumption include tolerances to account for human errors and style while conducting the tests. In a simulation study in \cite{Kubaisi2014}, these tolerances were explored concerning the power and energy demand. Different driver models were used to illustrate the difference a driver could introduce to these cycles. Their results suggest that an inexperienced driver in the WLTP cycle could increase the maximum power needed by 36\% compared to an idealized driver and some 5\% of energy consumption. A similar approach was taken in \cite{legg2022evaluating} where dynamometer tests are used to further tuning. 

This paper focuses on driver models that can utilize external cues such as road, traffic, etc. to end up in a driving behavior for the longitudinal direction, to assess energy consumption. Driver models for energy consumption studies are typically based on car-following models like the {\it intelligent driver model} (IDM) as in, e.g., \cite{hegde2021real}. In \cite{Canova2019} a car following model is enhanced to react to environmental stimuli other than the traffic as in conventional car following models such as traffic light and free driving. A switching concerning model between different operating conditions is proposed, where each mode has its mechanisms to generate the desired speed.  The standard IDM model typically models the combined behavior of the driver and the model in traffic simulations. In \cite{hegde2021real} a second layer is proposed to incorporate a more complete powertrain and vehicle model. This second layer is in the form of a simple PID tracking controller and the output of the enhanced driver model is regarded as the input of the PID controller. In \cite{schuermann2019model} a similar extended IDM model is developed that is capable of being calibrated using incorporated statistical maps. In \cite{Mcgordon2011} a driver model is proposed for fuel consumption studies. The fundamental idea is that the speed trace does not exist {\it a priori} but is rather generated as the vehicle moves forward. A driver interprets the surrounding environment and decides to generate a speed trace on the fly. Even though the results look promising in \cite{Mcgordon2011}, the report misses out on describing the actual model structure and mechanisms in detail.

The current paper expands on the idea of generating the speed trace on the fly from \cite{Mcgordon2011}, to fit into a framework previously developed by the authors for describing the use of vehicles, \cite{PetterssonStochastic} and \cite{ROMANO2021102878}. In this framework, a format was developed that can describe the relevant conditions of the vehicle w.r.t. to its energy consumption. Some of these conditions do not directly enter the equation of motion of the vehicle. A driver model is hence needed to interpret these conditions, such as upcoming curves, legal speed, etc, into the desired speed profile during a simulation. A driver model fitted to this framework was presented in \cite{pettersson2019operating}, following the structure of an operational and tactical part. The operational part of the driver model converts the desired speed into a pedal actuation and the tactical part interprets environmental cues into the desired speed. The objective of this paper is to parameterize and validate this model. 

The objective of the current paper is to check the validity of the assumptions made in the previously mentioned driver model. This includes the speed-choosing mechanism of driving in curves as well as the tuning of the operational controller part of the model. The model validation and parametrization are made against both driving simulator data recorded in a small simulator experiment as well as from log data recorded in real traffic operations.  Fidelity and accurate response are compared between the two data sources and discussed and in line with relevant literature, e.g.,~\cite{bittner2002road}. For the driving simulator data, known shortcomings of the environment are accounted for and data are not used for cases where the ecological validity is low, as opposed to \cite{chen2001identification} where the identified model includes uncertainties. 

\section{A driver model for energy efficiency}
In conventional driving cycles, the driver is often modeled as a controller that attempts to adjust the vehicle's speed to match a target profile \cite{AndersBengt,Jari3}. Accordingly, the inputs to the driver are the current values for the position $x$ along the road and the actual and desired speeds $v_x$ and $v_{\textnormal{d}}$, respectively, whilst the outputs are the vehicle actuation requests. A similar modeling approach, which substantially assimilates the driver to an \textit{operational module}, is sufficient when a target speed profile is postulated \textit{a priori} and independently of the environmental conditions. On the other hand, to translate the external stimuli coming from the surroundings into a desired speed signal, the operational model needs to be complemented by a \textit{tactical part}. Ideally, this should translate the information from the road into a speed input for the operational driver, based on an intuitive understanding of the physical phenomena governing the vehicle's longitudinal dynamics. Hence, the main idea behind the model proposed in this paper is to split the driver into a tactical and operational module, similarly to what was suggested by \cite{AndersEriksson}. 

\subsection{Tactical module}
The tactical part of the driver model interprets the surroundings based on simplified physical models and translates them into a desired speed input for the operational module. At each time step and vehicle position, the desired speed $v_{\textnormal{d}}$ is selected to maximize a comfort criterion while respecting the legal speed limit, according to
\begin{equation}\label{eq:dOCSpeeds}
v_{\textnormal{d}} = \min \Bigl\{v_{\textnormal{sign}}, v_{\textnormal{sign}}^\prime, v_\kappa, v_\kappa^\prime, v_{\textnormal{b}}, v_{\textnormal{b}}^\prime, v_{\textnormal{stop}}, v_{\textnormal{stop}}^\prime, v_{\textnormal{t}}, v_{\textnormal{t}}^\prime \Bigr\},
\end{equation}
where the speeds $v_{\textnormal{sign}}$, $v_\kappa$, $v_{\textnormal{b}}$ and $v_{\textnormal{stop}}$, $v_{\textnormal{t}}$ corresponds to the legal speed limit, road curvature $\kappa$, speed bumps and stops, and traffic density, respectively. Among these, only the first two speeds are considered in the present paper. It should be observed that these speeds are \textit{static}, meaning that they correlate with the local or current value for each involved physical quantity. In particular, inspired by kinematic considerations, the following empirical formula describes $v_\kappa$ as a function of the curvature $\kappa$:
\begin{equation}
    v_\kappa = \biggl( \frac{a_{y}^{\textnormal{max}}}{\kappa}\biggr)^\delta,
\label{eq:TacticalModel}
\end{equation}
where $a_{y}^{\textnormal{max}}$ is a comfort threshold for the lateral acceleration and $\delta$ is an exponent which may in turn depend upon the specific driving style. Specifically, the assumption of quasi-equilibrium maneuver performed at constant longitudinal speed yields $\delta = \frac{1}{2}$. However, albeit being legitimated from physical intuition, the value $\delta = \frac{1}{2}$ does not find confirmation from experimental evidence. In this context, it is worth noting that \autoref{eq:TacticalModel} may be restated more conveniently as
\begin{align}
    v_\kappa = \sqrt{\dfrac{\hat{a}_y^{\textnormal{max}}(\kappa)}{\kappa}}, && \textnormal{with} && \hat{a}_y^{\textnormal{max}}(\kappa) = a_y^{\textnormal{max}}\biggl( \dfrac{a_y^{\textnormal{max}}}{\kappa}\biggr)^{2\delta-1}. \label{eq:TacticalModel2}
\end{align}
The physical interpretation of \autoref{eq:TacticalModel2} is that, even in the case $\delta \not = \frac{1}{2}$, \autoref{eq:TacticalModel} may be still recast in the form of a quasi-equilibrium relationship between $v_\kappa$ and $\kappa$, provided that a curvature-sensitive comfort threshold $\hat{a}_y^{\textnormal{max}}(\kappa)$ for the lateral acceleration is considered in place of the constant value $a_y^{\textnormal{max}}$. 

Therefore, the kinematic relationships \autoref{eq:TacticalModel}, \autoref{eq:TacticalModel2} alternatively provide a static expression for the curvature speed $v_\kappa$. The static equations for $v_{\textnormal{b}}$, $v_{\textnormal{stop}}$ and $v_{\textnormal{t}}$ are instead not reported in the paper, and may be found in \cite{PetterssonStochastic,ROMANO2021102878}. 

For each speed appearing in \autoref{eq:dOCSpeeds}, a \textit{dynamic} version is also considered, which depends on the upcoming value for the related physical quantity. Denoting a generic static speed in \autoref{eq:dOCSpeeds} by $v_p$, the corresponding dynamic value $v_{p}^\prime$ is thus calculated as
\begin{equation}\label{eq:dynamicSpeed}
v_{p}^\prime = \sqrt{v_{p,i+1}-2 a_{x}^{\textnormal{max}}(x_{i+1}-x)},
\end{equation}
where $x$ is the current position, $x_{i+1}$ the next (discrete) position at which the speed $v_p$ changes value, and $a_x^{\textnormal{max}}$ is a longitudinal acceleration threshold. It should be clear that \autoref{eq:dOCSpeeds} and \autoref{eq:dynamicSpeed} will generate a continuous profile with piece-wise constant derivative. 

\subsection{Operational module}
The desired speed of the tactical module is actuated by the operational part of the driver model.  The difference between the desired and actual speeds is used to produce the pedal outputs $ a _ {\textnormal {p}} $ and $ b _ {\textnormal {p}} $, with $a_{\textnormal{p}}, b_{\textnormal{p}} \in [0, 1]$. In this paper, the operational part of the driver is modeled using a PID controller. Accordingly, the pedal positions $a_{\textnormal{p}}$ and $b_{\textnormal{p}}$ are be computed as
\begin{subequations}
\begin{align}
a_{\textnormal{p}} &= \left\{\begin{array}{lc}
    0, &   f_{\textnormal{PID}}(e) < 0 \\
    f_{\textnormal{PID}}(e), &  0 < f_{\textnormal{PID}}(e)< a_{\textnormal{p,max}}\\
     1, & 1 < f_{\textnormal{PID}}(e)
\end{array}\right.\\
-b_{\textnormal{p}} &= \left\{\begin{array}{lc}
     -1, & f_{\textnormal{PID}}(e)<-1\\
    f_{\textnormal{PID}}(e), &  -1 < f_{\textnormal{PID}}(e)< 0\\
     0, &  0<f_{\textnormal{PID}}(e)  
\end{array}\right.
\end{align}
\end{subequations}
where $e(t) = v_{\textnormal{d}}(t)-v_x(t)$ with $v_{\textnormal{d}}$ being the tactical speed, $v_x$ the vehicle speed and $f_{\textnormal{PID}}(\cdot)$ is the function describing the PID control law \cite{ROMANO2021102878}:
\begin{equation}
   f_\textnormal{PID}\bigl(e(t)\bigr) = K_{\textnormal{P}}e(t) + K_{\textnormal{I}}\int_{0}^t e(\tau)\dif\tau + K_{\textnormal{D}}\dod{e(t)}{t},
    \label{eq:OptDrvMdlPID}
\end{equation}
in which $K_{\textnormal{P}}$, $K_{\textnormal{I}}$ and $K_{\textnormal{D}}$ are the proportional, integral and derivative gains, respectively. It is worth emphasizing that the operational model described above neglects the effects associated with the driver's neuromuscular behavior. Such a simplification is rather popular when it comes to longitudinal applications like, for instance, lap-time simulations \cite{GermanStuff}. 

The implemented controller corresponds to the PID structure described above, with the addition of an anti-windup mechanism. Specifically, a conventional back-calculation anti-windup scheme is employed to prevent integrator windup when the throttle or brake command reaches its saturation limits. The method feeds back the difference between the saturated and unsaturated controller outputs through a proportional gain to the integrator, thereby preventing excessive growth of the integral state and ensuring a smooth recovery from saturation (see, e.g., \cite{aastrom2010feedback} for more details). 

\begin{figure}
\centering \includegraphics[width=1\linewidth]{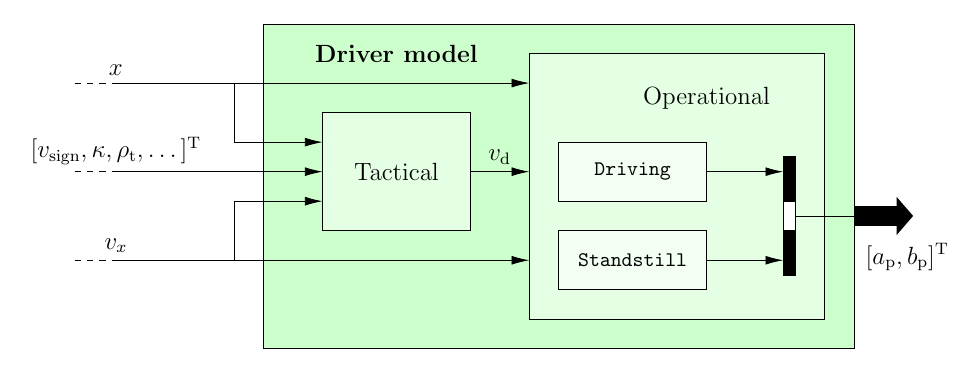}
\caption{Schematic representation of the driver model. The inputs are the vehicle's longitudinal speed and position $[v_x, x]^{\mathrm{T}}$ and the parameters which determine the speeds in \autoref{eq:dOCSpeeds}. The outputs are the acceleration and brake pedal positions $[a_{\textnormal{p}}, b_{\textnormal{p}}]^{\mathrm{T}}$.}
\label{fig:VehPropDriver}
\end{figure}

A pictorial depiction of the driver model, together with the input and output quantities, is shown in \autoref{fig:VehPropDriver}.

\section{Methods}
The objective of the paper is to parameterize and validate the previously proposed driver model using data recorded from a driving simulator study as well as from trucks in real traffic operation. This section gives a brief introduction to the data as well as to the methods used to parameterize and validate using this data.

\subsection{Data from driving simulator} 
\begin{figure*}
    \centering
    \includegraphics[width=0.8\textwidth]{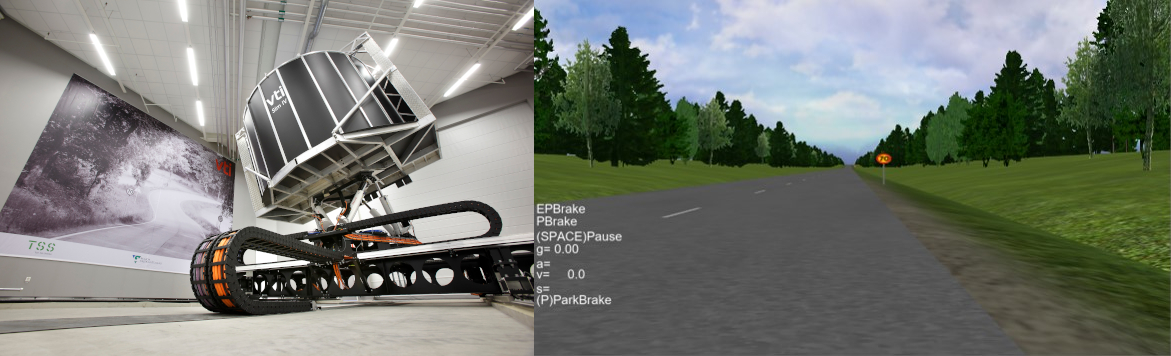}
    \caption{{\bf Left:} the motion platform of the simulator IV at VTI. {\bf Right:} an example view of the simulated environment in the study}
    \label{fig:simivenvironment}
\end{figure*}

A simple and small driving simulator study was conducted using 8 na\"ive professional truck drivers. The drivers filled in an informed consent form before the test. The aim of the study was to generate data that could be used to parameterize the driver model and to identify possible behaviors not captured by the present driver model. The driving simulator SimIV, \cite{jansson2014design} was used in the study, see \autoref{fig:simivenvironment}.
The experimental condition was a drive of some 10 minutes on a country road with a surrounding randomly generated forest. The driver was instructed to respect the speed limits and other traffic rules for the road and to behave as normally as possible compared to regular driving. The simulated vehicle (and the real cabin in the simulator) was a Volvo FH tractor with a semitrailer. The trailer was simulated to be loaded to the legal limit in the European Union (40 tons of gross combined weight). No other vehicles were present in the simulation. The driver could see the motion of the trailer in the rearview mirrors. All retarder functionality apart from the friction breaks was removed in the experiment, and the gear shifting was automatized. The test drivers were all instructed and informed about the vehicle and its configuration.

The intent was to use the recorded data to parameterize both the tactical and the operational part of the driver model. The simulated environment can be controlled in detail, which is utilized in the experimental setup. The road was set up such that the driver's behavior is a response to a single event and cue. Three different events were studied,
\begin{itemize}
    \item Road slope
    \item Speed limit change
    \item Road curvature
\end{itemize}
The events on the road were constructed to be isolated, e.g., when there was a slope, there were no speed limit changes and curvatures. Each event was separated by a road stretch of about 350 meters and with a nominal speed limit of 70 $\textnormal{km}\,\textnormal{h}^{-1}$. The order of the events was balanced such that all drivers experienced different event orders. 

For the event of road slope, 6 different slopes were used, three uphill and three downhill. For the speed limit change, 4 changes were implemented, two up and two down in speed and for the road curvature three left and three right turn radius according to \autoref{tab:simeventdata}.
\begin{table}[]
    \centering
    \begin{tabular}{ccc}
    \hline
        Road slope & Speed limit change & Road radius \\\hline
         $\pm5.5\%$ & $\pm 30 \textnormal{km}\,\textnormal{h}^{-1}$ & $250$ m \\
         $\pm3.5\%$ & $\pm 10 \textnormal{km}\,\textnormal{h}^{-1}$ & $175$ m \\
         $\pm1.5\%$ &  & $125$ m \\
         \hline
    \end{tabular}
    \caption{Simulator event data}
    \label{tab:simeventdata}
\end{table}

The speed limit and road curvature events were used to statistically understand the tactical part of the driver model. For this purpose,  a separation in time constants between the operational and tactical parts of the driver model is assumed. Hence, the effect of the operational part can be separated in the data and not explicitly considered in the analysis. Instead, an assumption is made that the statistical analysis is robust enough to cope with potential influence in the data from the operational part.

The road slopes and speed limit changes are used to parameterize the operational part of the driver model. The tactical part during these events is assumed to be trivial, i.e. constant set speed that the operational part tries to track. The analysis method is further described in the following subsection, including assumptions and problem formulations to obtain the parameters of the driver model. 

A general note should be made already here regarding using driving simulators for the purpose of tuning models. Generated data that for this purpose needs to be regarded from an absolute validity perspective. This would ideally require an ecological validity of the simulator environment that makes the drivers' responses and behaviors identical to the ones the driver would be having in an identical real traffic situation with a real vehicle. This validity is obviously not met in any driving simulator, independently of its technical complexity and quality. Driving simulator environments have some well-known weaknesses regarding their ecological validity. One major of these weaknesses is poor speed perception,  see, e.g.,~\cite{fischer2012evaluation}. 

A poor speed perception will influence the validity of the data used for the driver model and needs to be considered throughout the process of parameterizing and assessing the model. For example, actions have been taken for the data used when driving in the curves. The entrance speed into a curve is likely to suffer more from the simulator's poor speed perception due to a lack of cues leading up to the chosen speed, i.e. looming in the visual cue. This effect is of course present in real driving but more pronounced in a driving simulator. Hence, the entrance speed of the curve is not used in the analysis, but in the middle section. In the middle section of the curve, there are more cues active that will increase the driver's awareness of his/her speed, i.e. the lateral acceleration actuated by the motion platform of the simulator.


\subsection{Field data}



In this work, a database of log-data supplied by Volvo Trucks was used for the purpose of studying the speed adaption to curves. The database was created as part of the OCEAN project \cite{ocean}. 
The database contains data from 33 different heavy truck vehicles, where each truck traveled multiple routes of various distances. In total about 1000 logfiles were available, with each logfile containing signals from several sources, e.g., GPS altitude, latitude and longitude, the brake pedal position, etc. The data used in this study are the longitudinal vehicle speed, computed from wheel speed sensors, and the angular velocity of the vehicle around its vertical axis (yaw rate), telling how fast the truck is rotating left or right in the horizontal plane and used as a proxy for the value of the lateral (centripetal) acceleration of the truck. A steady curve yields a constant yaw rate while a sharp turn generates a rapidly changing yaw rate. 
 
The scope of this study is to assess the validity of the curve speed-radius relationship, \autoref{eq:TacticalModel}, recast into the expression of \autoref{eq:TacticalModel2}. Therefore, only the information collected along the curves travelled by the trucks is of interest. The curves were extracted from the database by analyzing every single route by means of a curve-detection algorithm which was developed in a previous study, see \cite{Maghsood2016.A1,Maghsood2016.A2}. Starting from the recorded logfile, the algorithm allowed obtaining the initial and final point of the curves, their length, i.e. the arc-length covered by the truck around the curve, as well as their direction (left or right). 

In order to obtain meaningful results, the curves were extracted and analyzed only from the routes corresponding to a travelled distance larger than 5 km. Moreover, the analysis of the measured longitudinal speed and lateral acceleration  profiles along the detected curves showed that both quantities are most stable in the central portion of the curve. Speed adjustments primarily occurred before entering and after exiting the curve, while the lateral acceleration remained approximately constant in the middle segment. Consequently, the curve analysis was performed over the central part of the curves (middle portion) since this was considered to be more representative than the one performed over its full length. 

\subsection{Identification of operational parameters}
The main idea to identify the parameters of the operational part of the driver model is to use the straight slope parts of the recorded driving simulator data. The assumptions made on the conditions in which the data was acquired is that the drivers intend to keep a constant speed, dependent on the legal (signed) speed. With a change in road grade, the driver experiences an external excitation, or disturbance, in the shape of a slope change.

Data from the speed limit changes could potentially be used in a similar fashion as the data from the road slopes. However, initial tests using this data revealed that the drivers' decision on when to change the speed varies across the test persons and has a substantial influence on the system identification. The descision when to adjust a new legal speed is also not covered by the driver model. Hence, it was decided that this data was not used for the purpose of identifying the operational part of the driver model. 

The identification process of the PID control parameters described above is set up to minimize the response of the simulated system compared to the recorded driving simulator data. The additional proportional gain of the wind-up is not included in the identification, since it does not contribute significantly to the model behavior. The driver model is in a closed loop with the vehicle and represents a challenging identification problem as the input of the driver (speed error) is not independent of the output (the brake and acceleration pedal positions). Instead, the problem is considered as a conventional grey box identification problem where the input is the set speed and the output is the pedals' positions, see, e.g., \cite{book:Ljung1999} for more details on the fundamentals of close loop identification. In this approach, the input is the road slope, which is independent of the pedal positions. 

The driver model act in a closed loop with the vehicle, which is perfectly known in both the parametrization process as well as in the data recordings. A vehicle dynamics model is used to generate the response of the driver's pedal positions, to generate the motion feed to the motion platform and computer graphics in the driving simulator. In the identification process, the very same model is used in a close loop with the driver model. In the following, a brief presentation of the longitudinal part of the used vehicle dynamics model is presented.

The longitudinal response of the vehicle is in general described by a simple model of the powertrain and the dynamics of motion. It is in principle a few state models of the form,
\begin{subequations}
\begin{align}
\dot{v}_x &=\frac{T_q(\omega_{\textnormal{e}})a_{\textnormal{p}} n_{\textnormal{g}}}{mr}-g\sin \theta-\frac{F_rr+F_{\textnormal{air}}(v_x)+K_{\textnormal{brake}}b_{\textnormal{p}}}{m}, \\
\dot{\omega}_{\textnormal{e}} &=G(p)\Big(\frac{v_x}{rn_{\textnormal{g}}}\Big),
\end{align}
\end{subequations}
where $a_{\textnormal p},b_{\textnormal p}\in[0,1]$ are the acceleration and brake pedals positions, $v_x$ longitudinal speed of the vehicle, $n_{\textnormal G}$ is the gear ratio of the current gear, $G(\cdot)$ is a transfer function (linear dynamics) from road wheel to engine speed and $T_q$ is the engine map dependent on the engine speed $\omega_{\textnormal e}$.

The identification is then formulated as an optimization problem where the objective is to minimize the pedal responses and the speed error,
\begin{equation}
\begin{split}
\min_{K_{\mathrm P},K_{\mathrm I},K_{\mathrm D},
      v_d}
\int_{t_{\mathrm{sim}}}
\Big(
    &(a_p^{\mathrm{meas}}(t)-a_p(t))^2 \\
    +&(b_p^{\mathrm{meas}}(t)-b_p(t))^2 \\
    +&\frac{1}{20^2}(v_d-v_x(t))^2
\Big)\,\dif t
\end{split}
\label{eqn:optimization}
\end{equation}
where the superscript $(\cdot)^{\textnormal{meas}}$ indicates data from the logfiles of the simulator and $v_d$ is the set speed of the tactical driver model. It should be noted that this reference speed, from the tactical part of the driver model, is not a given quantity, and the optimization is very sensitive to this reference. An assumption in the model is that this is a constant quantity as long as there are no external stimuli, for example, signed speed or upcoming curvature, etc. Hence, the set speed is part of the optimization problem to prevent a biased estimate of the controller parameters through an erroneous set speed. To scale the influence of the set speed in the objective function, a factor is introduced. For all events, the legal (signed) speed was $70 \, \textnormal{km}\,\textnormal{h}^{-1} \approx 20\,\textnormal{m}\,\textnormal{s}^{-1}$. Hence, this factor was also used to scale all three quantities in the range between zero and one approximately.

\section{Results}
This section presents the results obtained from the parametric of the driver model using both driving simulator data as well as logfiles measured from truck vehicles in real-world operations. The results are presented separately for the operational and tactical parts. 

\subsection{Operational part}
\begin{figure}[h!] 
\centering
  \includegraphics[width=0.49\textwidth]{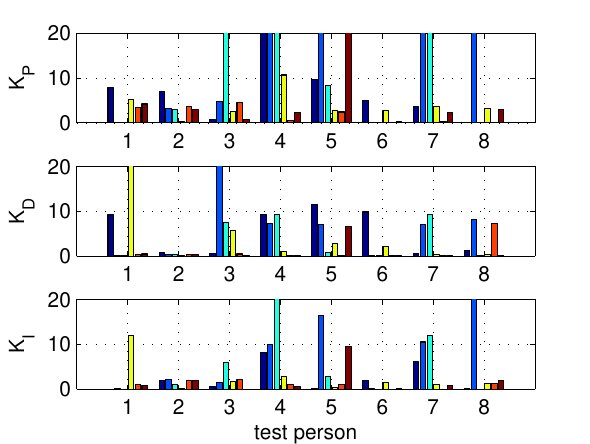}
  \includegraphics[width=0.49\textwidth]{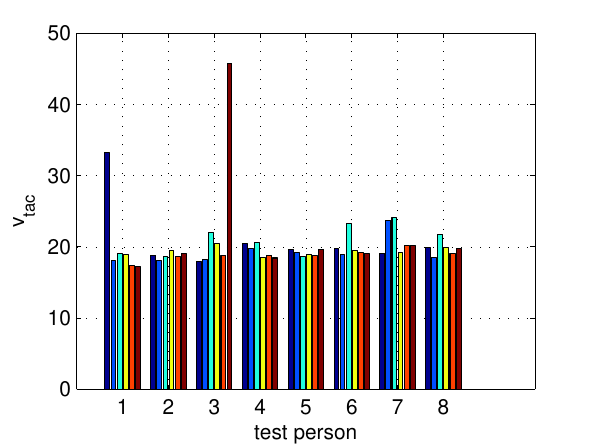}
\caption{Result of the optimization of the operational model parameters (PID controller parameters) ({\bf left}). Each event and driver is a separate optimization. the tactical speed in the optimization of operational model parameters. Signed speed was $70~\textnormal{km}\,\textnormal{h}^{-1} \approx 19~\textnormal{m}\,\textnormal{s}^2$ ({\bf right}).}
\label{fig:ResultOptimalUnsort}
\end{figure}

The parameters of the operational model were identified using the optimization formulation given in \autoref{eqn:optimization}. Each event was regarded separately, and an optimization problem was solved across drivers and events, generating six sets of parameters for each driver of the operational model. This approach, as opposed to, for example, obtaining one set of parameters per driver, enables a study on how the characteristics of the event might have an influence on the driver. The complete set of parameters for all drivers in the driving simulator data set is presented in \autoref{fig:ResultOptimalUnsort}.

The optimization problem \autoref{eqn:optimization} was complemented with box constraints on the model parameters to more easily detect non-converging problems. Different starting points were tested to check whether a local optimum of the solution was obtained. The optimization routine was the MATLAB \verb=fmincon=, which uses the standard interior-point method without providing an analytical derivative.

In \autoref{fig:ResultOptimalUnsort}, it can be observed that the identified model parameters exhibit considerable variability, both across drivers and across individual events. Moreover, the vertical axis is truncated; consequently, bars reaching the upper limit represent parameter values that exceed the displayed range. Part of this variability, particularly the exceptionally large parameter estimates, may be explained by violations of the assumption of a constant tactical speed. In such cases, the identified parameters become difficult to interpret unambiguously.
\begin{figure}[h!] 
\centering
  \includegraphics[width=0.49\textwidth]{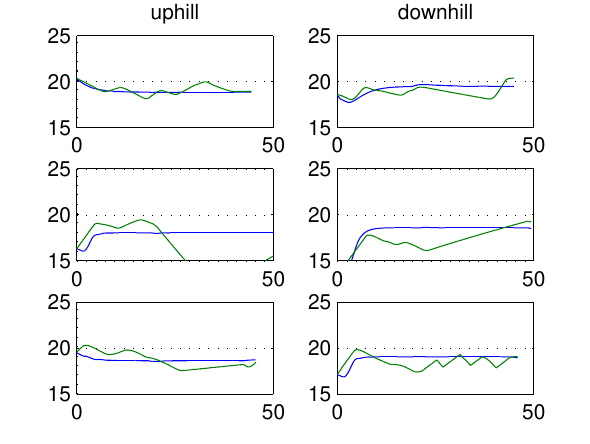}
  \includegraphics[width=0.49\textwidth]{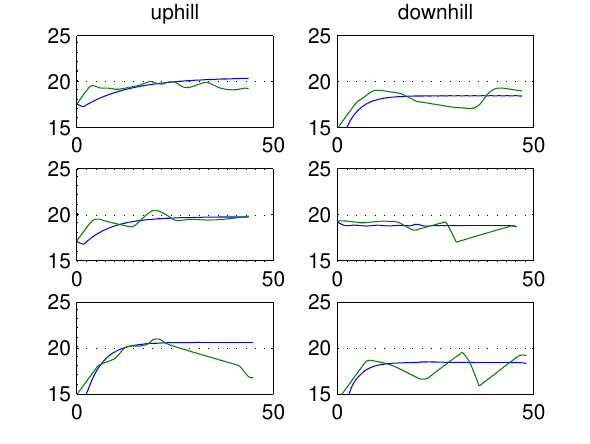}
\caption{The speed profile of driver number 2 (left) and 4 (right) with the recorded trace in green and the simulated, with optimal parameters, in blue.}
\label{fig:ResultOptimalDriver2and4}
\end{figure}

An example of the violation of the assumption of constant tactical speed can be seen in \autoref{fig:ResultOptimalDriver2and4}, where the speed traces of drivers 2 and 4 are plotted together with the speed trace obtained from the simulation. Both drivers maintain a constant speed for most of the events, whereas driver 4 exhibits more pronounced deviations. However, the objective of the optimization is also to minimize the model output error (pedal positions), making the explanation of violating the constant speed less obvious.

\begin{figure}[h!] 
\centering
  \includegraphics[width=0.49\textwidth]{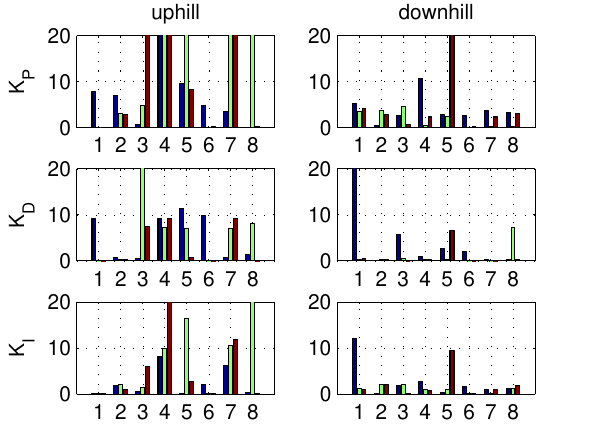}
\caption{Same results as in \autoref{fig:ResultOptimalUnsort}, but separated between uphill and downhill events.}
\label{fig:ResultOptimalSort}
\end{figure}

When the identified parameter sets are separated into uphill and downhill events (\autoref{fig:ResultOptimalSort}), a clear pattern emerges: the parameter estimates associated with downhill events generally lie within the expected range, whereas those corresponding to uphill events frequently approach the imposed parameter bounds. This suggests that driver behavior during uphill propulsion is more difficult to capture with the proposed model and that vehicle power limitations increase the sensitivity of the parameter identification. In addition, gear selection plays a more prominent role during uphill driving, introducing further complexity that is not explicitly represented in the current model. 

It is a commonly known fact that speed perception is poor in humans when there are sensory cues (vibrations, changing environment, etc.). This is even stronger in a simulated environment like a driving simulator. An idea could be to model a behavior similar to this as a zero-speed perception, except when the real speed and the tactical speed are deviating significantly. By replacing the PID controller part of the model $f_\textnormal{PID}(e)$ in \autoref{eq:OptDrvMdlPID} with
\begin{equation}
f_\textnormal{SW}(e)=\left\{\begin{array}{ll}
    K_pe & e\le-e_\textnormal{SW}   \\
    f_\textnormal{SW}(e^-) & -e_\textnormal{SW}<e<e_\textnormal{SW}\\
    K_pe & e_\textnormal{SW}<e
    \end{array}\right.
    \label{eq:SwitchOpDrvMdl}
\end{equation}
where $f_\textnormal{SW}(e^-)$ indicates function value in previous time instant and $e_\textnormal{SW}$ is a switching threshold and $e=v_\textnormal{tac}-v_x$. With this model, the driver will act as a proportional controller whenever the error between the speed and the tactical speed is above the threshold. When the error decrease below the threshold, the pedal position is held until the error has grown large again. 

\begin{figure}[h!] 
\centering
  \includegraphics[width=0.49\textwidth]{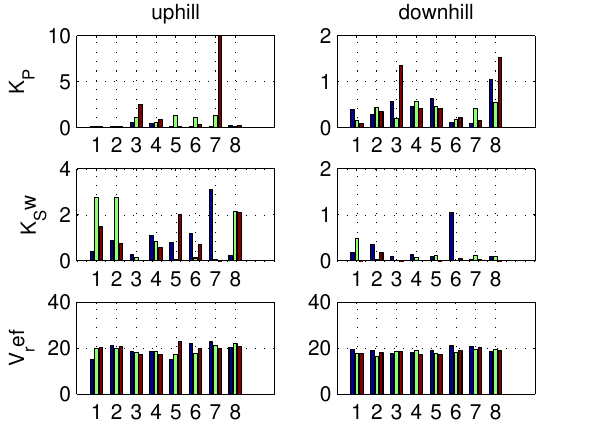}
\caption{Same results as in \autoref{fig:ResultOptimalUnsort}, but for the switched proportional controller and separated between uphill and downhill events.}
\label{fig:ResultOptimalSwitchedSort}
\end{figure}

The identifcation results of replacing the conventional PID controller with \autoref{eq:OptDrvMdlPID} is depict in \autoref{fig:ResultOptimalSwitchedSort}.

\subsection{Tactical part}
\label{sec:ANOVA}

The speed adaptation due to curves is here investigated, both for driving simulator data and for field data. For the driving simulator data, the acceleration in speed adaptation due to a change in legal speed is investigated.

The model for speed adaptation due to curves, see~\autoref{eq:TacticalModel} and \autoref{eq:TacticalModel2}, is here re-formulated as
curve speed, $v_\kappa$, versus curve radius, $R$, according to
\begin{equation}
    v_\kappa = v_{0}  \left( \frac{R}{R_{0}} \right)^{\delta},
    \label{eq:Speed_vs_Radius}
\end{equation}
where the model parameters are the exponent $\delta$, and the speed $v_0$, corresponding to the speed in a reference curve with radius $R_0$; in this study the reference curve radius is chosen to $R_0=250$~m. Note that the logarithm the above equation becomes a linear model:
\begin{equation}
    \ln \left(v_\kappa\right) = \ln v_{0} + \delta \left( \ln R - \ln R_{0} \right),
    \label{eq:Speed_vs_Radius2}
\end{equation}
which will be used for estimating parameters from data.

\subsubsection{Curve analysis for driving simulator data}
\label{sec:SimulatorCurveAnalysis}

\begin{figure}
	\centering
    \includegraphics[width=1\linewidth]{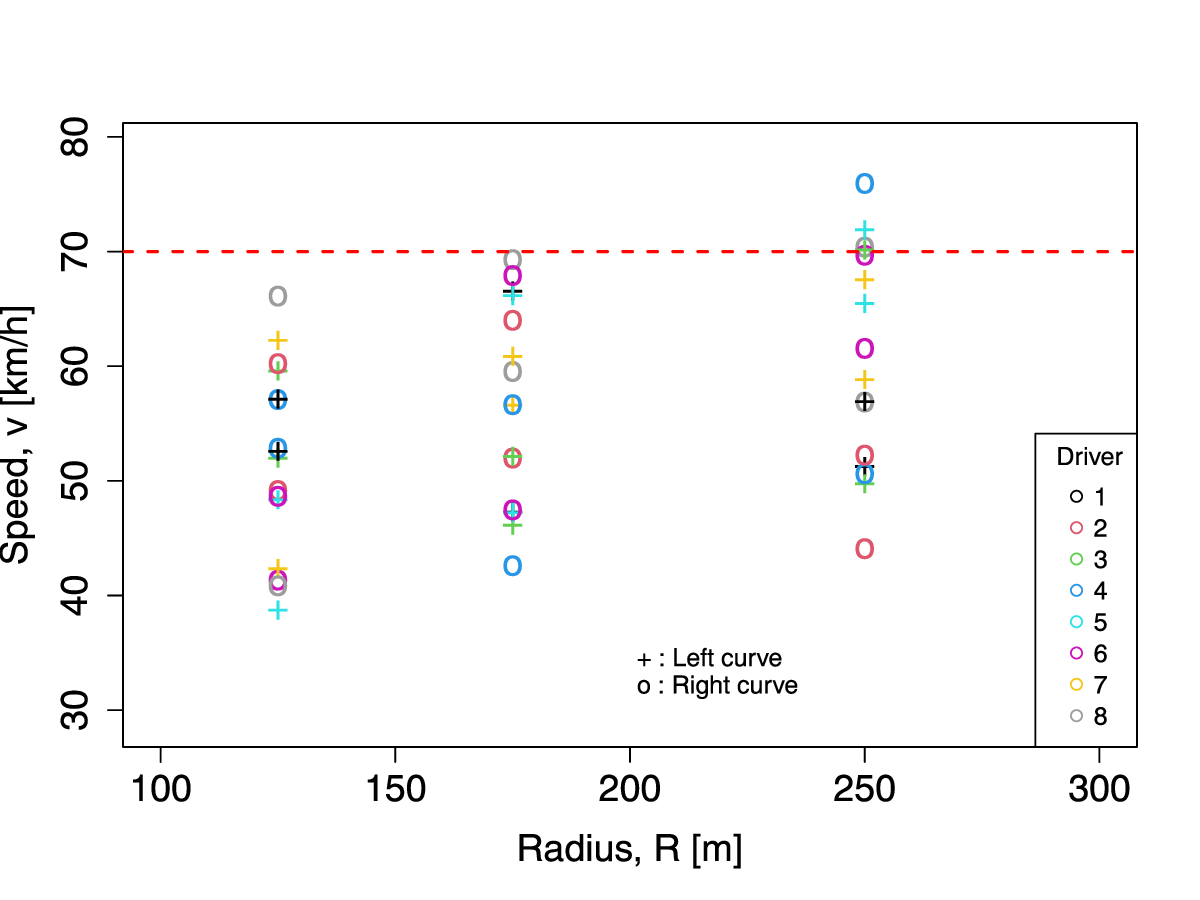}
	\caption{Data from driving simulator: Speed in curves as function of curve radius ($R$ = 125, 175, 250 m). The colours represent the 8 different drivers, and the symbol represents curve direction, Left (+), Right (o). The dashed line shows the speed limit.}	
    \label{fig:SpeedCurvesSimulatior}
\end{figure}

Each of the eight drivers encounters six curves with three different curve
radius, $R = 125, 175, 250$~m, and each curve radius is driven in both
left and right curve directions. The speed limit is 70 km/h for all
encountered curves. The speed in the curves has been extracted from the
speed signal in the driving simulator data, and is presented in \autoref{fig:SpeedCurvesSimulatior}. The speed in each curve has been calculated as the average speed over
the part of the curve with maximum curvature.

A statistical analysis was performed, evaluating the influence of the speed in curves on the curve radius, the curve direction, and the driver. 
For this purpose, a linear model was formulated by studying the logarithm of the speed model, \autoref{eq:Speed_vs_Radius2}, namely
\begin{equation}
    \ln\left( v_{kij} \right) = \ln v_{0} + \delta \left( \ln R_{k} - \ln R_{0} \right) + \alpha_{i} + \ \beta_{j} + e_{kij},
\end{equation}
where $v_{kij}$ is the measured speed in a curve with radius $R_{k}$ for driver $i$ in curve direction $j$ (1='Left', 2='Right'), and $e_{kij}$ is a normally distributed random error. The categorical variables $\alpha_{i}$ with $i\in \{1,\ldots,8\}$ and $\beta_{j}$ with $j\in \{1,2\}$ represent the driver effect and the curve direction effect, respectively. 
An analysis of variance (ANOVA) was performed for the linear model, and the resulting ANOVA table
is presented in \autoref{tab:ANOVA_simulator_speed}. 
The conclusion from the ANOVA is that both radius and driver show very strong effects and are statistically significant ($p$-values less than $10^{-10}$), whereas the effect of curve direction is not
statistically significant.

\begin{table*}[]
\caption{ANOVA table for speed in curves for driving simulator data.
}
    \label{tab:ANOVA_simulator_speed}
    \centering
    \begin{tabular}{lrrrrrl}
        \multicolumn{7}{l}{
        Model considered: $\ln\left( v_{kij} \right) =  \ln v_{0} + \delta \left( \ln R_{k} - \ln R_{0} \right) + \alpha_{i} + \beta_{j} + e_{kij}$} \\
        \hline
        \textit{Source of Variation} & \textit{Df} & \textit{Sum Sq} & \textit{Mean Sq} & \textit{$F$-value} & \textit{$p$-value} & \\
        \hline
        Radius, $\ln(R_k)$        &  1 & 0.20526 & 0.205265 & 87.6916 &   $2.05\cdot 10^{-11}$ &  *** \\
        Driver, $\alpha_{i}$  &  7 & 1.07727 & 0.153896 & 65.7463 &  $< 2.2\cdot 10^{-16}$ &  *** \\
        Curve direction (L/R), $\beta_{j}$    &  1 & 0.00021 & 0.000207 &  0.0882 &   0.768    \\
        Residuals, $e_{k}$   & 38 & 0.08895 & 0.002341                       \\
        \hline
        \multicolumn{7}{l}{Signif.\ codes: 0 ‘***’ 0.001 ‘**’ 0.01 ‘*’ 0.05 ‘.’ 0.1 ‘ ’ 1.}
    \end{tabular}
\end{table*}

\begin{figure}
	\centering
    \includegraphics[width=1\linewidth]{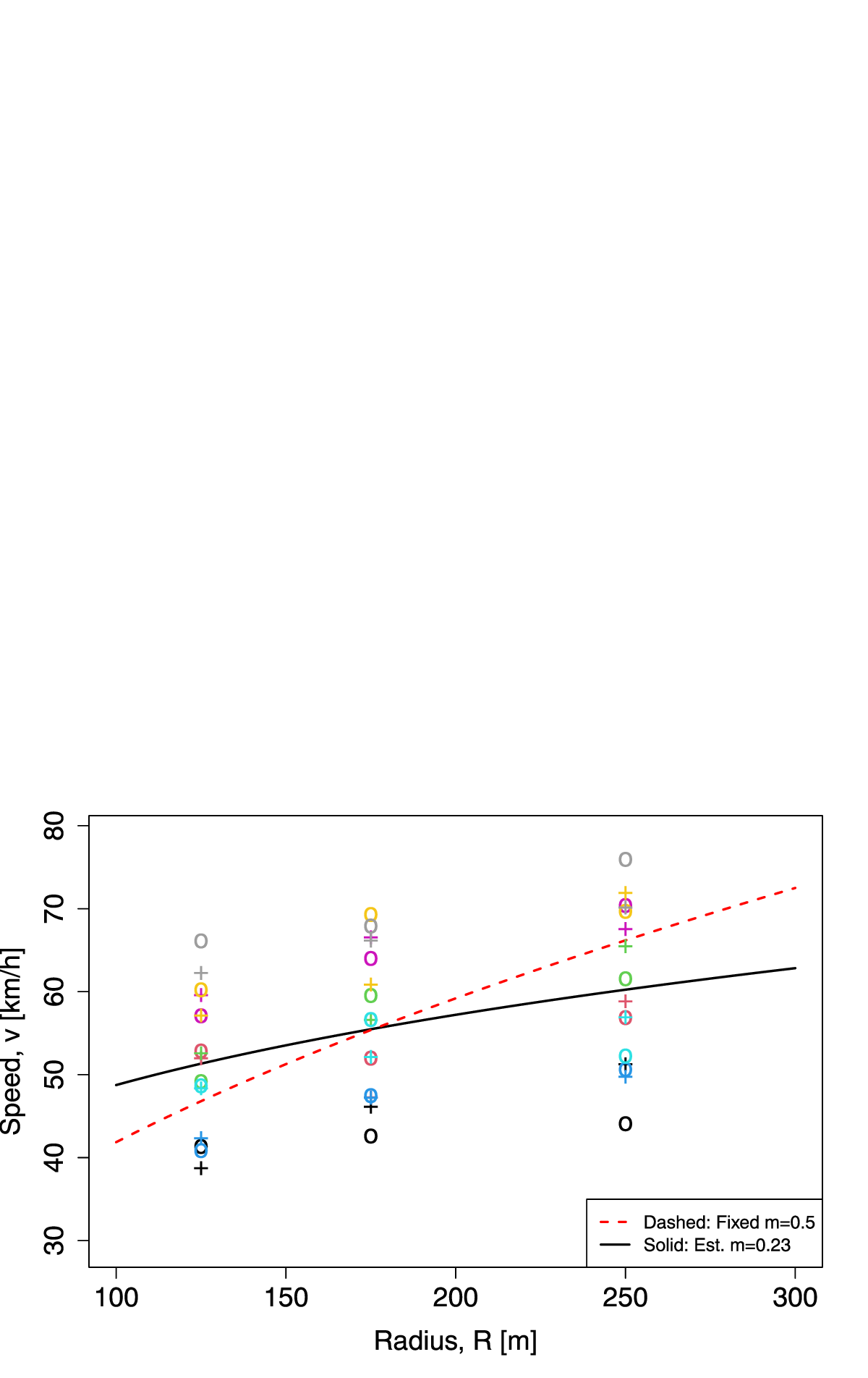}
	\caption{Curve speed model as function of curve radius.}	
	\label{fig:EstCurveSpeedSimulator}
\end{figure}

Therefore, in the continued analysis, the linear model is reduced by excluding the term corresponding to the curve direction, resulting in the linear model
\begin{equation}
    \ln\left( v_{kij} \right) = \ln v_{0} + \delta \left( \ln R_{k} - \ln R_{0} \right) + \alpha_{i} + e_{kij}.
\end{equation}
The estimated parameters are presented in \autoref{tab:EstPar_simulator_speed}, and the resulting speed model is shown in \autoref{fig:EstCurveSpeedSimulator}. 
The speed in the reference curve with radius $R_0=250$~m (which is also the largest radius in the driving simulator study) is estimated to 60~km/h, which is lower than the speed limit of 70~km/h. This strongly indicates that the drivers have adopted the speed to the curve radius.
The exponent $\delta$ reflects how much the driver adapts the speed to the curve radius; a higher value represents a stronger adaptation. For the driving simulator data the exponent is estimated to $\delta=0.23$ with 95\% confidence interval of $(0.18 ; 0.28)$, see \autoref{tab:EstPar_simulator_speed}.
Often an exponent $\delta = 0.5$ is suggested, corresponding to a fixed maximum lateral comfort acceleration in curves, see \autoref{eq:TacticalModel} and the discussion on the model. 
The hypothesis of $\delta = 0.5$ is rejected by this driving simulator study, since the confidence interval for $m$ does not cover the value 0.5.
However, it should be noted that the speed may be perceived differently in a real driving environment compared to the artificial environment in the driving simulator. Therefore, the results should be treated with care. 

Further, it can be observed that the driver variation in curve speed is quite large (16\%), whereas the residual variation is much smaller (about 5\%). 
The conclusion is that, for the driving simulator study, the majority of the speed variation in a curve with fixed radius is contributed to the between driver effect.


\begin{table*}[]
\caption{Estimated parameters for speed in curves for driving simulator data.}
    \label{tab:EstPar_simulator_speed}
    \centering
    \begin{tabular}[]{lll}
        \hline
        Parameter & Estimated exponent & Fixed exponent
        \\
        \hline
        Exponent, $\delta$ & 0.23 (0.18 ; 0.28) & 0.5 (fixed)\\
        Speed at radius \(R_{0} = 250\ m\), \(v_{0}\) & 60 km/h & 66
        km/h \\
        Driver variation, \(\text{std}(\alpha_{i})\) & 16\% &  16\%
         \\
        Residual standard deviation, \(\text{std}(e_{kij})\) & 4.8\% &
        9.6\% \\
        \hline
    \end{tabular}
\end{table*}

For illustration purposes, the model with fixed exponent $\delta = 0.5$ is evaluated with estimated parameters presented in \autoref{tab:EstPar_simulator_speed}, and the estimated model is illustrated by the dashed line in \autoref{fig:EstCurveSpeedSimulator}. It can be observed that the unexplained variation, represented by the residual, increases, which is due to the lack of fit to the model for data at high and low curve radius.


\subsubsection{Curve analysis for field data}
\label{sec:CurveAnalysis}



In this section, curve speeds extracted from customer field data will be evaluated with respect to the adaptation of the speed to the curve radius.
Each one of the curves detected by the curve detection algorithm was analyzed as follows. For the $k$:th curve, the curve speed, $v_k$, was derived from the experimentally determined speed profile as the average speed over the central part of the curve. In similar fashion, the radius of the curve, $R_k$, was derived as the average of the radius along the central part of the curve.

\begin{figure*}[h!] 
\centering
  \centering
  \includegraphics[scale=0.83]{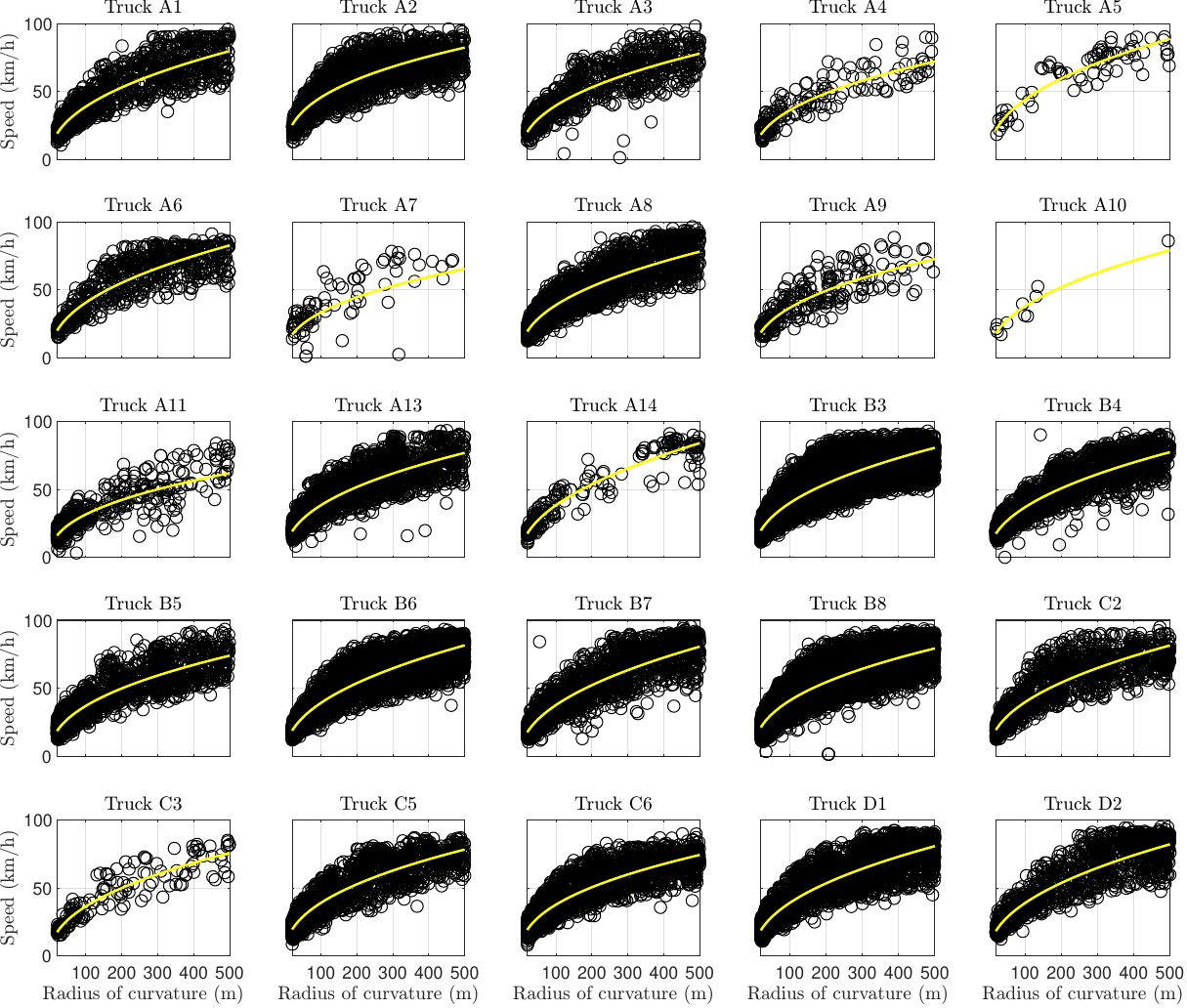}
\caption{Field data: Speed-radius diagrams of the curves extracted for all trucks of  
the Volvo Trucks field dataset and obtained regression models (solid lines). }
\label{fig:RadiusSpeedAcceleration}
\end{figure*}

\autoref{fig:RadiusSpeedAcceleration} shows the  speed-radius diagrams resulting from plotting the so-obtained values for $v_k$ and $R_k$. The diagrams are organized by truck, with only the trucks travelling routes longer than 5~km being included in the analysis. As visible from the figure, a plateau is reached in the longitudinal speed for radii larger than about 500 m. This plateau corresponds to the speed limits or is a consequence of speed limitations inside the truck. On the other hand, it is believed that curves with very small radius may correspond to situations like parking. For this reason, it was decided to perform regression analysis only on the curves corresponding to radii comprised between 20~m and 500~m. 

A regression analysis was carried out as a preliminary step prior to the ANOVA. The regression model \eqref{eq:Speed_vs_Radius2} was applied to all the routes traveled by each truck in the database. 
In \autoref{fig:RadiusSpeedAcceleration} the resulting regression models are superposed to the experimental data points, while \autoref{tab:RegressionModels} gives the list of the parameters resulting from regression. 

\begin{table*}[h!] 
\caption{Parameters of the regression model - tactical model.}
\centering
\resizebox{\textwidth}{!}{\begin{tabular}{@{}llllllllllllllllllllllllll@{}}
\hline
Truck & A1 & A2 & A3 & A4 & A5 & A6 & A7 & A8 & A9 & A10 & A11 & A13 & A14 & B3 & B4 & B5 & B6 & B7 & B8 & C2 & C3 & C5 & C6 & D1 & D2 \\
\hline
$m$ (-) & 0.45 & 0.37 & 0.42 & 0.43 & 0.44 & 0.44 & 0.42 & 0.43 & 0.42 & 0.46 & 0.42 & 0.43 & 0.48 & 0.43 & 0.46 & 0.43 & 0.46 & 0.47 & 0.41 & 0.45 & 0.45 & 0.43 & 0.42 & 0.45 & 0.47   \\
$v_{0}$ ($km/h$) & 58 & 64 & 58 & 54 & 65 & 61 & 49 & 58 & 54 & 58 & 47 & 57 & 60 & 60 & 56 & 55 & 59 & 58 & 60 & 60 & 55 & 58 & 56 & 59 & 60  \\
\hline
\end{tabular}}
\label{tab:RegressionModels}
\end{table*}

An ANOVA was performed to extract information for the formulation of the tactical part of the driver model (\autoref{eq:TacticalModel}). Generally speaking, the purpose of an ANOVA is to determine whether data from several groups have a common mean, that is, to find out whether different groups of the input variable have different effects on the response variable. In this framework, the ANOVA was performed to quantify the effect of different drivers (trucks), and curve radius and direction, in the formulation of the tactical part of the driver model. 

As for the driving simulator study, a linear model was formulated by studying the logarithm of the speed model, \autoref{eq:Speed_vs_Radius2}, namely
\begin{equation}
    \ln\left( v_{kij} \right) = \ln v_{0} + \delta \left( \ln R_{k} - \ln R_{0} \right) + \alpha_{i_k} + \beta_{j_k} + e_{k},
\end{equation}
where $v_{kij}$ is the measured speed in the $k$:th curve with radius $R_{k}$, for truck $i_k$ and curve direction $j_k$ (1='Left', 2='Right'); $e_{k}$ is a normally distributed random error. The categorical variables $\alpha_{i}$ with $i\in \{1,\ldots,n_{truck}\}$ and $\beta_{j}$ with $j\in \{1,2\}$ represent the truck (or driver) effect and the curve direction effect, respectively. 

An ANOVA was performed for the linear model, and the resulting ANOVA table
is presented in \autoref{tab:ANOVA_logdata_speed}. 
The conclusion from the ANOVA is that both radius and driver show very strong effects and are statistically significant ($p$-values less than $10^{-15}$), whereas the effect of curve direction is not as strong and is barely statistically significant with p-value of 0.012. 

\begin{table*}[]
\caption{ANOVA table for speed in curves for field data.
}
    \label{tab:ANOVA_logdata_speed}
    \centering
    \begin{tabular}{lrrrrrl}
        \multicolumn{7}{l}{
        Model considered: $\ln\left( v_{kij} \right) =  \ln v_{0} + \delta \left( \ln R_{k} - \ln R_{0} \right) + \alpha_{i} + \beta_{j} + e_{kij}$} \\
        \hline
        \textit{Source of Variation} & \textit{Df} & \textit{Sum Sq} & \textit{Mean Sq} & \textit{$F$-value} & \textit{$p$-value} & \\
        \hline
        Radius, $\ln(R_k)$     &  1 & 9867.8 & 9867.8 & $3.16\cdot 10^{-5}$ &   $< 2\cdot 10^{-16}$ &  *** \\
        Driver, $\alpha_{i}$  &  24 & 60.1 & 2.5 & 80.2  &  $< 2\cdot 10^{-16}$ &  *** \\
        Curve direction (L/R), $\beta_{j}$  &  1 & 0.2 & 0.2 &  6.27 &   0.0123 &  *   \\
        Residuals, $e_{kij}$   & 49283 & 1540.1 & 0.0313                       \\
        \hline
        \multicolumn{7}{l}{Signif.\ codes: 0 ‘***’ 0.001 ‘**’ 0.01 ‘*’ 0.05 ‘.’ 0.1 ‘ ’ 1.}
    \end{tabular}
\end{table*}


The estimated parameters are presented in \autoref{tab:EstPar_logdata_speed}.
The speed in the reference curve with radius $R_0=250$~m is estimated to 54.5~km/h, and the data demonstrate that the drivers have adopted the speed to the curve radius.
Recall that a higher value of the exponent $\delta$ implies a stronger speed adaptation by the driver to the curve radius. For the field data the exponent is estimated to $\delta=0.46$ with 95\% confidence interval of $(0.458 ; 0.462)$, see \autoref{tab:EstPar_logdata_speed}.
This is slightly smaller than the exponent $\delta = 0.5$, which corresponds to a fixed maximum lateral comfort acceleration in curves, see \autoref{eq:TacticalModel} and the corresponding discussion. 
The hypothesis of $\delta = 0.5$ is rejected by the analysis of the field data, since the confidence interval for $\delta$ does not cover the value 0.5; however, the estimated value of 0.46 is quite close to 0.5. 
Further, it can be observed that the driver variation in curve speed is substantial (5.6\%), however the residual variation is much larger (about 18\%). The curve direction effect was found to be small, with about 0.4\% higher speed in left-oriented curves compared to right-oriented curves.


\begin{table*}[]
\caption{Estimated parameters for speed in curves for field data.}
    \label{tab:EstPar_logdata_speed}
    \centering
    \begin{tabular}[]{lll}
        \hline
        Parameter & Estimated exponent & Fixed exponent\\
        \hline
        Exponent, $\delta$ & 0.460 (0.458 ; 0.462) & 0.5 (fixed) \\
        Speed at radius \(R_{0} = 250\ m\), \(v_{0}\) & 54.5 km/h & 56.2 km/h \\
        Driver variation, \(\text{std}(\alpha_{i})\) & 5.6\% &  4.5\%   \\
        Curve direction, \(\beta_{j}\) & $\pm 0.2\%$ &  $\pm 0.3\%$   \\
        Residual standard deviation, \(\text{std}(e_{kij})\) & 17.7\% &  18.1\%  \\
        \hline
    \end{tabular}
\end{table*}

For illustration purposes, the model with fixed exponent $\delta = 0.5$ is evaluated with estimated parameters presented in \autoref{tab:EstPar_logdata_speed}.
It can be observed that the unexplained variation, represented by the residual, increases, which is expected due to the slightly worse fit to the data when fixing the exponent to $\delta = 0.5$.

The conclusion from the field data study is that a large part of the variability of curve speed can be attributed to the between driver (truck) variation, which is estimated to 5.6\%, and that the effect of curve direction (left or right) is much smaller, about $\pm 0.2\%$, with slightly higher speed in left oriented curves than right oriented curves. However, the unexplained residual variation is the largest contribution to the variability in curve speed, about 18\%. The residual variation covers the within driver variation, i.e., there is a variability in speed for the encounter curves of a specific driver. The reason could also be factors that has not been taken into account in the current analysis, e.g., surrounding traffic, visibility in curves or road quality, but there is also expected to be a random variation within a specific driver.


\subsubsection{Speed changes -- driving simulator}
\label{sec:SimulatorSpeedChanges}


When a driver encounters a speed limit sign, its speed is expected to be adjusted to the legal speed. The characteristic acceleration and deceleration associated with such speed transitions are examined in this section.

\begin{figure}
	\centering
    \includegraphics[width=1\linewidth]{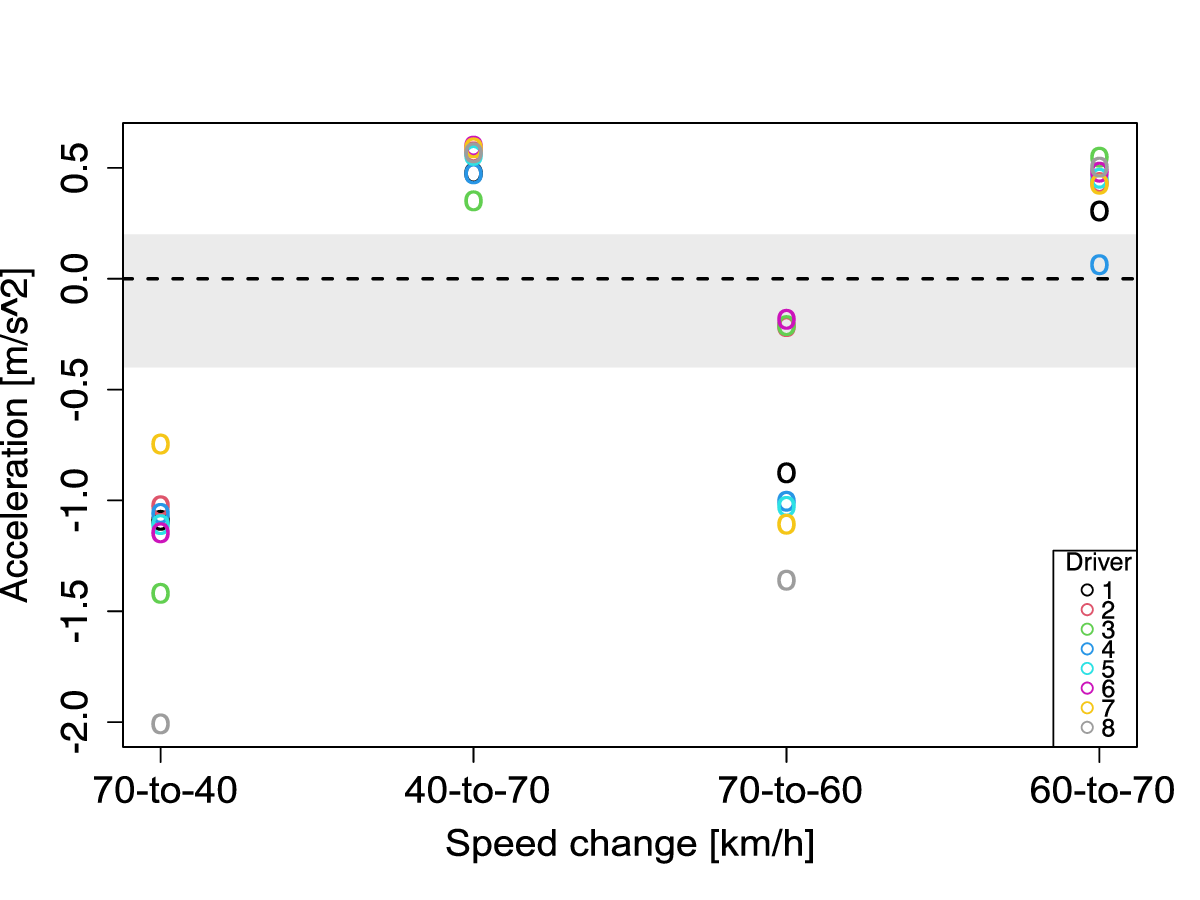}
	\caption{Accelerations/retardations due to changes in legal speed for speed changes 70-to-40, 40-to-70, 70-to-60, 60-to-70 km/h. The colours represent the 8 different drivers.}
	\label{fig:SpeedChanges_data}
\end{figure}

The acceleration at speed changes due to change of legal speed has been estimated from the driving simulator data and is presented in \autoref{fig:SpeedChanges_data}. The acceleration has been calculated as the average acceleration over a period with largest acceleration. For the small speed changes (70-to-60 km/h and 60-to-70 km/h) some data points show almost zero acceleration, since the vehicle already had the desired speed and thus no speed change was necessary. Therefore, only the data points corresponding to an acceleration larger than 0.2 $\textnormal{m}\,\textnormal{s}^2$ or smaller than -0.4 $\textnormal{m}\,\textnormal{s}^2$ (outside the gray shaded area) has been included in the analysis.

A model for the measured acceleration $a_k$ is proposed according to 
\begin{equation} 
a_k=a_0 \cdot \exp(e_k), 
\end{equation}
where the parameter $a_0$ is the median acceleration, and $e_k$ is a normally distributed random error, $e_k \sim N(0,\sigma^2)$. Note that this model represents a log-normal distribution.

First, each speed change category is evaluated separately. Note that there is only one replicate for each driver for each speed change category. Therefore, the diver effect cannot be estimated, and thus the potential diver effect will here be included in random error. 

The estimated parameters of median acceleration $a_0$, standard deviation $\sigma$, for each speed change category is presented in \autoref{tab:EstPar_simulator_speedChange}. It can be observed that the retardation is larger than the acceleration, which seems reasonable since the acceleration is limited by the engine power but the retardation is limited mainly by the brake efficiency. It can also be observed that the acceleration/retardation is slightly smaller for smaller speed changes compared to larger speed changes.

\begin{table}[]
\caption{Estimated parameters for acceleration/retardation in speed changes due to change in legal speed based on driving simulator data.}
    \label{tab:EstPar_simulator_speedChange}
    \centering
    \begin{tabular}[]{lrrrr}
        \hline
        Parameter & 70-to-40 & 40-to-70 & 70-to-60 &  60-to-70 \\
        \hline
        Median, $a_0$ [$\textnormal{m}\,\textnormal{s}^2$]    & -1.16 & 0.52 & -1.06 & 0.44 \\
        SD,  $\sigma$ [-]       & 0.28 & 0.18 & 0.16 & 0.18  \\
        \hline
    \end{tabular}
\end{table}




\section{Discussion}

The present study aimed at assessing the validity of a driver model designed to operate within an operating cycle framework, with a particular focus on the separation between tactical and operational components. The results obtained from both driving simulator data and field data provide several insights into the strengths and limitations of the proposed modeling approach.

For the operational part, the identification procedure revealed a significant variability in the PID parameters across both drivers and driving events. This variability indicates that the assumption of a single, fixed set of controller parameters may be insufficient to describe driver behaviour under varying conditions. In particular, the distinction between uphill and downhill events highlighted the influence of vehicle dynamics constraints, such as limited propulsion capability and gear selection, on the resulting driver response. In these cases, the identified parameters often approached the imposed bounds, suggesting that the underlying model structure struggles to fully capture the driver behaviour when the vehicle operates close to its physical limits.

Furthermore, the assumption of a constant desired speed during certain events was shown to be only partially valid. Deviations from this assumption contribute to ambiguities in the parameter identification process. The proposed alternative model based on a switching proportional controller suggests that drivers may not continuously regulate speed in a linear feedback manner, but rather act intermittently when the deviation from the desired speed exceeds a certain threshold. This behaviour is consistent with known limitations in human speed perception, particularly in simulated environments.

Regarding the tactical part, the analysis of curve demonstrates that the proposed speed–radius relationship provides a reasonable description of driver behaviour. However, the exponent $\delta$ deviates from the theoretically expected value of $0.5$, especially in the driving simulator study where a significantly lower value was obtained. This suggests that drivers do not strictly follow a constant lateral acceleration strategy, but instead adopt a more conservative or perception-driven approach, particularly in low-fidelity environments.

The comparison between simulator and field data highlights important differences. While the simulator data exhibits a stronger between-driver effect and lower residual variability, the field data shows a larger unexplained variability. This can be attributed to the presence of additional influencing factors in real traffic, such as surrounding traffic, road conditions, and visibility, which are not explicitly accounted for in the model. The closer agreement of the exponent $\delta$ to $0.5$ in the field data suggests that real-world driving behaviour aligns more closely with physical constraints than simulator-based behaviour.

Overall, the results confirm that the proposed driver model captures the main trends in driver behaviour relevant for energy consumption simulations. At the same time, the findings indicate that both the operational and tactical components would benefit from extensions that explicitly account for variability in driver behaviour and environmental conditions.

\section{Conclusions}

This paper investigated the parameterization and validation of a driver model intended for use within an operating cycle framework for energy consumption simulations of road truck vehicles. The model is based on a decomposition into a tactical component, generating a desired speed from environmental inputs, and an operational component, responsible for tracking this speed.

The operational component, implemented as a PID controller, was calibrated using driving simulator data. The results show that the model can reproduce observed speed and pedal responses under certain conditions, but also reveal a considerable variability in the identified parameters across drivers and events. In particular, limitations were observed in situations involving uphill driving, where vehicle constraints and gear selection influence the driver behaviour. These findings suggest that a fixed-parameter controller may not be sufficient to capture the full range of driver responses, and that simplified alternative formulations, such as switching control strategies, may provide a more robust description.

The tactical component was evaluated through the analysis of speed adaptation to road curvature using both driving simulator data and field data. The proposed power-law relationship between speed and curve radius was supported by the data. However, the exponent governing this relationship differed between simulator and real-world conditions, with simulator data exhibiting a weaker adaptation to curvature. This highlights the influence of perception-related effects in simulated environments and underlines the importance of validating models against real-world data.

The comparison between simulator and field data further showed that real-world driving behaviour exhibits larger variability, likely due to unmodelled environmental factors such as traffic interactions and road conditions. Despite these differences, the model was able to capture the general trends in speed adaptation and provides a physically interpretable framework for incorporating driver behaviour into energy consumption simulations.

In conclusion, the proposed driver model constitutes a viable and transparent approach for simulating driver behaviour in operating cycle-based energy assessments. Future work should focus on extending the model to account for additional environmental inputs, incorporating stochastic variability, and improving the representation of driver perception and decision-making processes.

\section*{ACKNOWLEDGMENT}
The authors whish to acknowledge the project OCEAN, “Operating cycle energy management”, Dnr 2013-006720 and the FFI program at the Swedish Energy Agency for financial support. 

\bibliographystyle{IEEEtran}
\bibliography{refs}

\begin{IEEEbiography}[{\includegraphics[width=1in,height=1.25in,clip,keepaspectratio]{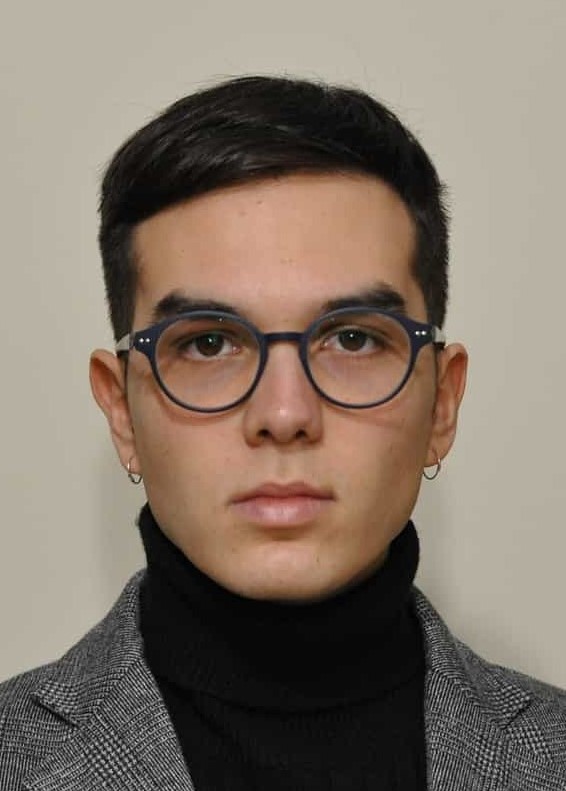}}]{LUIGI ROMANO } (Member, IEEE) received his M.Sc. in mechanical engineering from the University of Naples “Federico II”, Italy, in 2018. In 2023, he obtained his Ph.D. in machine and vehicle systems from Chalmers University of Technology, Gothenburg, Sweden, and a second M.Sc. in applied mathematics from the University of Gothenburg. Since 2024, Luigi has been affiliated with the Vehicular Systems group at the Department of Electrical Engineering, Linköping University, where he is a VR Postdoctoral Fellow funded by the Swedish Research Council. His primary research interests include vehicle dynamics and automotive control. In 2024, Luigi was awarded the AIMETA Junior Prize from the Italian Association of Theoretical and Applied Mechanics (AIMETA) for his research on tire and vehicle dynamics. Luigi is the author of the book \emph{Advanced Brush Tyre Modelling} (Springer, 2022).
\end{IEEEbiography}

\begin{IEEEbiography}[{\includegraphics[width=1in,height=1.25in,clip,keepaspectratio]{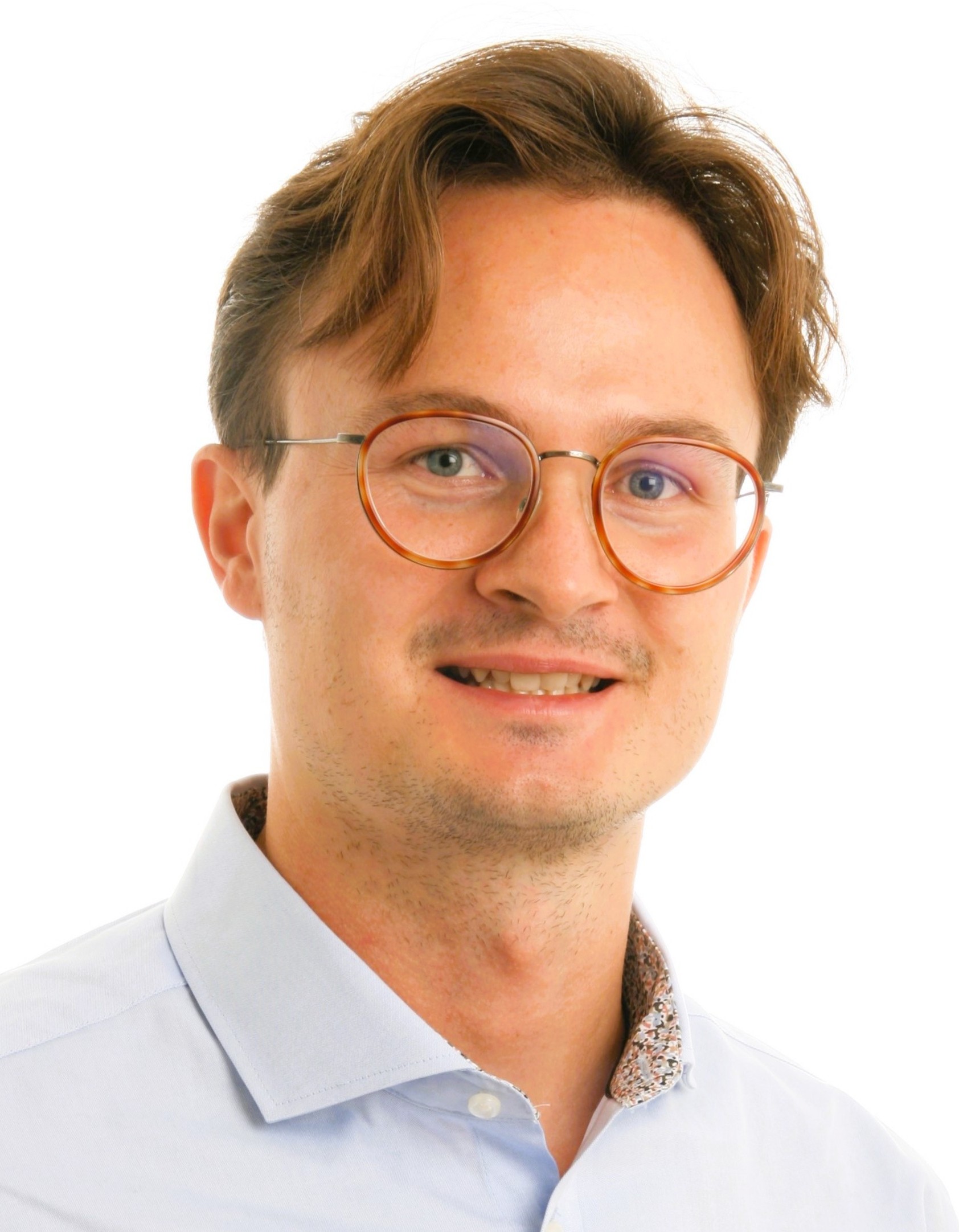}}]{MICHELE GODIO } received a double master’s degree in Civil and Structural Engineering from Politecnico di Milano, Italy, and École des Ponts ParisTech, France, as well as an advanced master’s degree in Mechanics of Materials and a Ph.D. degree in Mechanics of Materials and Structures from École des Ponts ParisTech. In October 2019, he joined the Department of Applied Mechanics at RISE Research Institutes of Sweden as a Researcher, following a four-year postdoctoral appointment at the Earthquake Engineering and Structural Dynamics Laboratory of the École Polytechnique Fédérale de Lausanne (EPFL), Switzerland. 

His research interests include the dynamics of materials and structures subject to extreme loadings, investigated through analytical, numerical, and experimental approaches developed in collaboration with numerous international partners.
\end{IEEEbiography}

\begin{IEEEbiography}[{\includegraphics[width=1in,height=1.25in,clip,keepaspectratio]{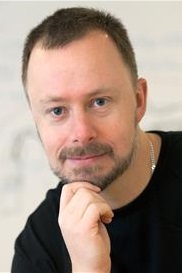}}]{FREDRIK BRUZELIUS }  received the M.S. degree in applied mathematics from Link\"oping University, Sweden, in 1999 and a PhD in control theory and Docent in vehicle dynamics from Chalmers, Sweden in 2004 and 2014 respectively. He is currently  an associate professor at Chalmers in vehicle dynamics.

From 2008 to 2022, he was a Senior Researcher in vehicle dynamics at the Swedish National Road \& Transport research institute (VTI). At VTI worked with issues connected to vehicle dynamics such as tire dynamics, modeling and testing and driving simulators. Prior to that, he was at Volvo Technology for some four years. 

Fredrik's research interest is broad within vehicle dynamics, and include modeling and control aspects of vehicles motion and energy consumption.
\end{IEEEbiography}

\begin{IEEEbiography}[{\includegraphics[width=1in,height=1.25in,clip,keepaspectratio]{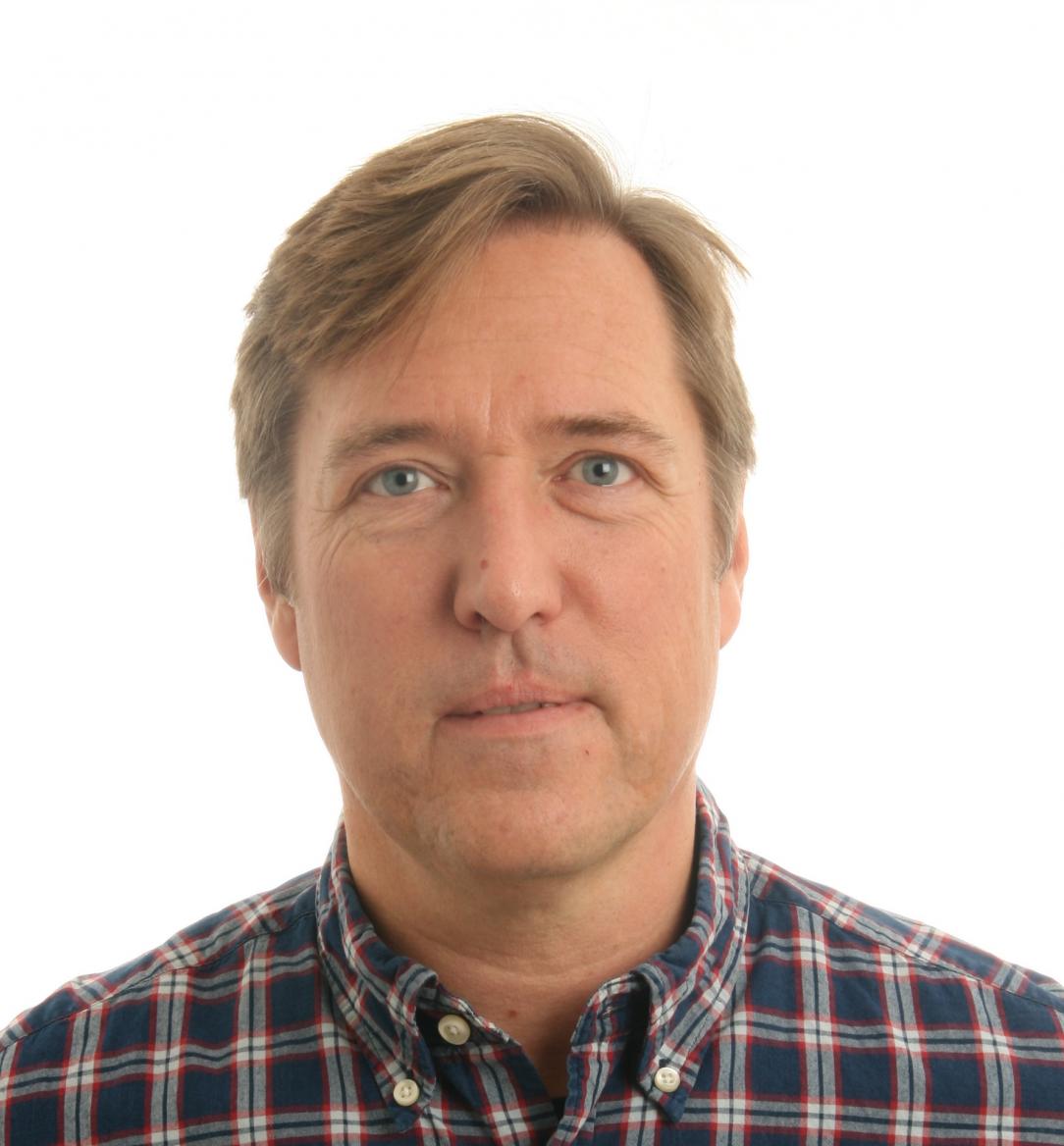}}]{P\"{A}R JOHANNESSON } received his Ph.D. degree in Mathematical Statistics from the Lund Institute of Technology, Lund, Sweden, in 1999, with a thesis on statistical load analysis for fatigue. He had a Postdoctoral position in mathematical statistics, at Chalmers University of Technology and PSA Peugeot Citro{\"e}n in France. He is also Docent in Mathematical Statistics at Chalmers University of Technology. He is currently a Senior Researcher in mechanical reliability, fatigue, and load analysis with RISE Research Institutes of Sweden, Gothenburg, Sweden. 

He has more than 30 years of experience in research and development in fatigue and load analysis with a focus on engineering statistics and reliability in industrial applications. He has published more than 25 articles in international journals. He is a Co-Editor of the handbook Guide to Load Analysis for Durability in Vehicle Engineering
\end{IEEEbiography}

\end{document}